# Inkjet printing-based volumetric display projecting multiple full-colour 2D patterns

Ryuji Hirayama[1,2,*], Tomotaka Suzuki[1], Tomoyoshi Shimobaba[1], Atsushi Shiraki[3], Makoto Naruse[4], Hirotaka Nakayama[5], Takashi Kakue[1] and Tomoyoshi Ito[1]

[1]Graduate School of Engineering, Chiba University, 1-33 Yayoi-cho, Inage-ku, Chiba 263-8522, Japan

[2]Research Fellow of the Japan Society for the Promotion of Science, 5-3-1 Kojimachi, Chiyoda-ku, Tokyo 102-0083, Japan

[3]Institute of Management and Information Technologies, Chiba University, 1-33 Yayoi-cho, Inage-ku, Chiba 263-8522, Japan

[4]Network System Research Institute, National Institute of Information and Communications Technology, 4-2-1 Nukui-kita, Koganei, Tokyo 184-8795, Japan

[5]Center for Computational Astrophysics, National Astronomical Observatory of Japan, 2-21-1 Osawa, Mitaka, Tokyo 181-8588, Japan

[*] hirayama@chiba-u.jp

In this study, a method to construct a full-colour volumetric display is presented using a commercially available inkjet printer. Photoreactive luminescence materials are minutely and automatically printed as the volume elements, and volumetric displays are constructed with high resolution using easy-to-fabricate means that exploit inkjet printing technologies. The results experimentally demonstrate the first prototype of an inkjet printing-based volumetric display composed of multiple layers of transparent films that yield a full-colour three-dimensional (3D) image. Moreover, we propose a design algorithm with 3D structures that provide *multiple different* 2D full-colour patterns when viewed from different directions and experimentally demonstrates prototypes. It is considered that these types of 3D volumetric structures and their fabrication methods based on widely deployed existing printing technologies can be utilised as novel information display devices and systems, including digital signage, media art, entertainment and security.

## Introduction

Volumetric displays directly render a three-dimensional (3D) image onto the true volume physical space [1–2]. Each volume element (voxel) of a 3D object is physically present at the required location and thus a natural visual perception is afforded from the surrounding. A variety of volumetric displays are intensively studied to achieve next-generation human–computer interaction and other applications [3–12].

In previous studies, an algorithm was proposed for the design of 3D structures, which projects *multiple* 2D patterns in different directions [13–15]. As shown in Fig. 1a, a 3D glass structure designed by the algorithm provides multiple 2D images independently to each of the viewpoints. As opposed to conventional 3D structures demonstrated in previous studies [16, 17], the algorithm in the present study provides grayscale images and projection directions that can be configured more flexibly. In addition, a multi-colour volumetric display based on the 3D arrangement of photoreactive luminescence materials was developed [18]. A volumetric display that projected three multi-colour 2D patterns was demonstrated by arranging two types of quantum dots that emitted red and green lights. The optical excitation of the voxels in the volumetric display resolves the issue of occlusion as opposed to conventional methods that necessitate electrical wiring, which significantly degrades the quality of 3D images.

However, all the manufacturing processes related to the aforementioned optically excited volumetric display were performed *manually*, i.e., the voxels were manually cut into a cubic shape and manually assembled to construct the volume. Accordingly, the volumetric display device reported in a previous study [18] comprised only $8 \times 8 \times 8$ voxels each with 5 mm side-lengths. This leads to difficulties in increasing the display resolution and representing full-colour images.

In this study, a manufacturing method is proposed with a high resolution volumetric display based on inkjet printing technology, which enables the minute and automatic location of photoreactive luminescence materials on appropriate places. Figure 1b shows the concept of the proposed method. The proposed volumetric display is composed of several layers of transparent films, and 2D patterns of fluorescent ink are printed on each of the films. The printed fluorescent materials arranged in a 3D layout are excited by external light irradiation (e.g. ultraviolet light), each excited fluorescent material, on returning to the ground state, emits light, which together form a 3D image. In this method, a massive amount of voxels and a full-colour representation can be achieved at a low cost using the widely available inkjet printers by adjusting the ratio of three fluorescent inks emitting primary colours (red, green and blue).

Furthermore, an algorithm proposed in a previous study [13] is extended to design a 3D structure that projected multiple full-colour 2D patterns, as shown in Fig. 1c. As described below, the study succeeded in experimentally demonstrating 3D structures projecting three and four full-colour patterns. Each of the patterns can only be reconstructed from a designated viewpoint according to the proposed volumetric display method based on inkjet printing.

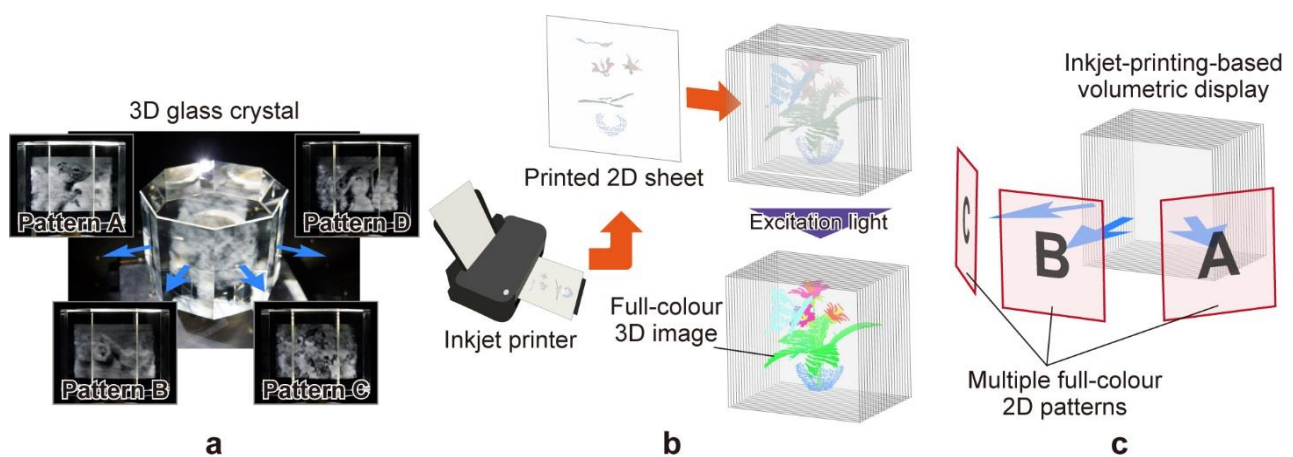

**Figure 1.** Concept of the study. (a) 3D glass structure projecting four monochromatic patterns in different directions. (b) Volume construction method enabling high resolution. (c) Inkjet printing-based volumetric display projecting multiple

full-colour 2D patterns.

# Results

## Full-colour 3D image representation

First, a prototype of the inkjet printing-based volumetric display is presented to demonstrate its ability to represent an arbitrary full-colour 3D image. Figure 2a shows the overview of the prototype. Fluorescent inks (SO-KEN Inc., Trick Print Ink) are used as photoreactive luminescence materials that mainly comprise europium complex (red), β-quinophthalone (green) and coumarin dyes (blue). An inkjet printer (SO-KEN Inc., 'TPW-105PB') prints the inks on 0.1 mm thick polyester transparent films (Folex imaging Inc., 'BG-32') with a maximum resolution of 5,760 × 1,440 dpi. The rendering space of the 3D image corresponds to 35 mm × 35 mm × 12.5 mm. Twenty layers of the printed films are stacked at 0.5 mm intervals.

In addition, 3D figures of flowers and butterflies comprising 51,767 full-colour points are used as a source for the 3D objects. Figure 2b shows twenty cross-sectional images of the 3D objects. Each layer includes 300 × 300 pixels and is printed on the films. In order to excite the printed fluorescent ink, an ultraviolet light peaked at 365 nm (AS ONE, LUV-4) is irradiated on the volumetric display in the perpendicular direction. Figure 2c shows the 3D objects rendered by a computer simulation viewed from different perspectives. Figure 2d presents images of an experimentally fabricated volumetric display obtained from different viewpoints. In particular, $\theta$ in Figs. 2c and 2d depicts a horizontal angle between the perpendicular to the films and the viewing direction. The highest quality of the 3D image is obtained when it is viewed in a direction perpendicular to the films ($\theta = 0°$). Although the images obtained from the diagonal directions ($\theta = \pm 30°$) are blurred, it is confirmed that the volumetric display appropriately represents motion parallax.

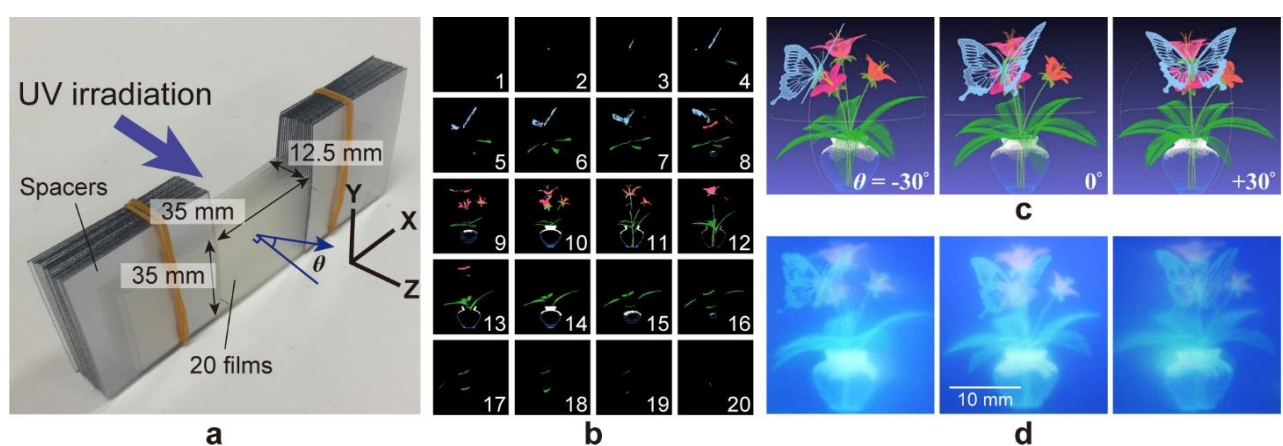

**Figure 2.** Prototype of the volumetric display based on inkjet printing. (a) Overview under natural light. (b) Cross-sectional images printed on the films. (c) Original 3D image represented with computer graphics from three different viewpoints. (d) Images of the volumetric display when excited by ultraviolet light.

## 3D structure projecting multiple full-colour 2D patterns

Next, a 3D structure projecting three different full-colour patterns in different directions is designed, and hereafter the patterns are referred to as patterns **A**, **B** and **C**. The concept of the algorithm presented in a previous study [13] is extended, and the extended algorithm is used to determine the voxel values of the full-colour 3D structure (please refer to the Method section for details). The original patterns comprised $512 \times 512$ pixels with three channels (red, green and blue), and each of the channels can represent 256 gradations. As shown in Fig. 1c, pattern **B** is set such that it is projected in the direction perpendicular to the films. The projection directions of the other two patterns (**A** and **C**) correspond to ±30° around the Y-axis. Figure 3a shows twenty cross-sectional images of the designed 3D structure when three full-colour images, as shown in Fig. 3b (right: **A**, middle: **B** and left: **C**), are used as the original patterns.

Simulations are performed to confirm the successful projection of original 2D patterns from the designed 3D structure. It should be noted that the simulation did not involve the effects of the light absorption by the films or inks since the primary objective involved confirming the validity of the design method. Figure 3c shows the three different patterns projected from the designed volumetric display. Deteriorations caused by other patterns (cross talks) are shown. However, the projected patterns are recognised as the original patterns, as shown in Fig. 3b. It is confirmed that the proposed algorithm [13] is extended to full-colour representations.

Twenty films with printed cross-sectional images are stacked to create a prototype of the volumetric display. The images in Fig. 3d correspond to the images of the volumetric display observed from different viewpoints. In order to decrease the effects due to photo absorption of the ultraviolet excitation light by the films (for details please refer to the experimental evaluation section), the volumetric display is irradiated by two ultraviolet lights placed on the top and the bottom of the display. The observed patterns are not as clear as the ones predicted in the simulations. Nevertheless, the three full-colour patterns are recognised from each viewpoint (see Supplementary Video S1).

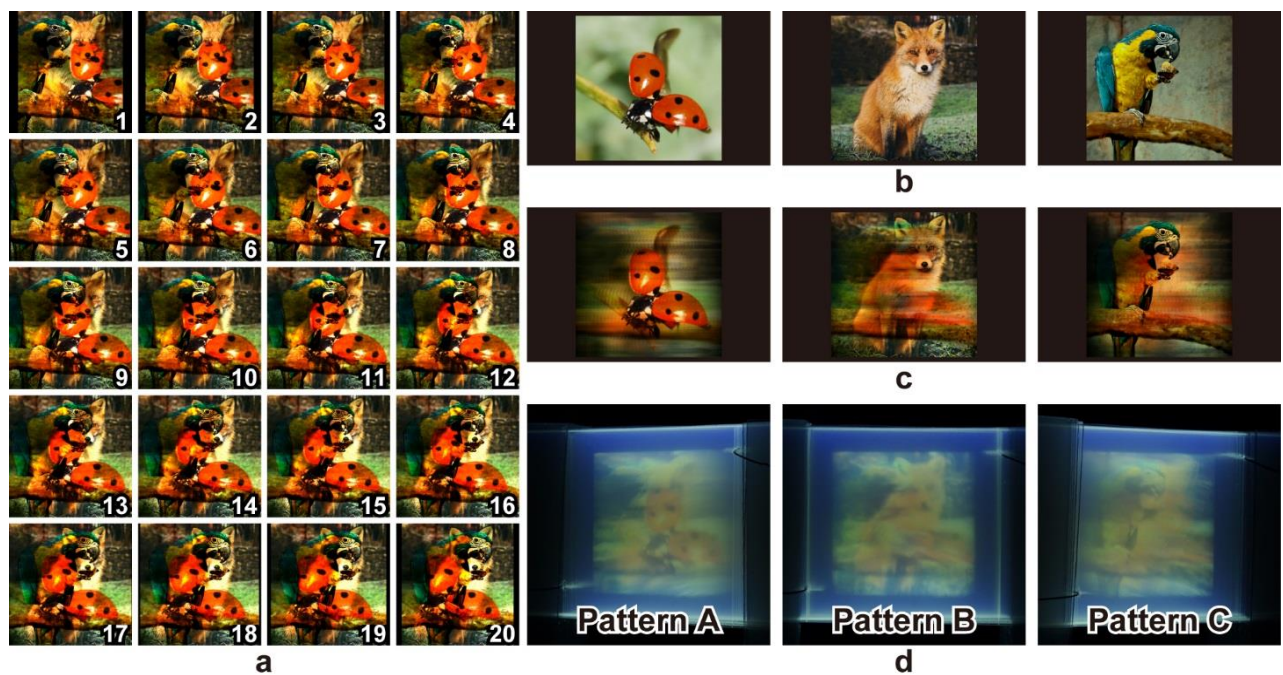

**Figure 3.** Prototype of the volumetric display projecting three full-colour patterns. (a) Cross-sectional images printed on the films. (b) Original images recorded on the volume. (c) Projected patterns from different viewpoints of the 3D structure in the simulation. (d) Projected patterns of the volumetric display (see also Supplementary Video S1). The photographs used as the original images can be found at http://free-photos.gatag.net/?s=201304180900 (pattern A),

https://unsplash.com/photos/HQqIOc8oYro (patterns B) and https://unsplash.com/photos/clkYWlgOHIQ (pattern C). All of them are licensed under the Creative Commons Public domain (https://creativecommons.org/publicdomain/zero/1.0/deed.en).

The projection axes can actually be configured freely. As shown in Fig. 5a, a prototype of the volumetric display projecting *four* patterns is created to experimentally demonstrate such a characteristic. In the experiment, the Z-axis is rotated ±20° around both the X- and Y-axes to derive the projection axes. Figure 5b shows the simulation results of the projected patterns from the display. Degradation is observed in the contrast of the projected patterns when compared with that in Fig. 3c. However, four different patterns are recognised from each viewpoint. The images in Fig. 5c correspond to the images of the experimentally fabricated volumetric display observed from different viewpoints. Although the images are not clear when compared with that of the simulations, the four different patterns can be successfully observed.

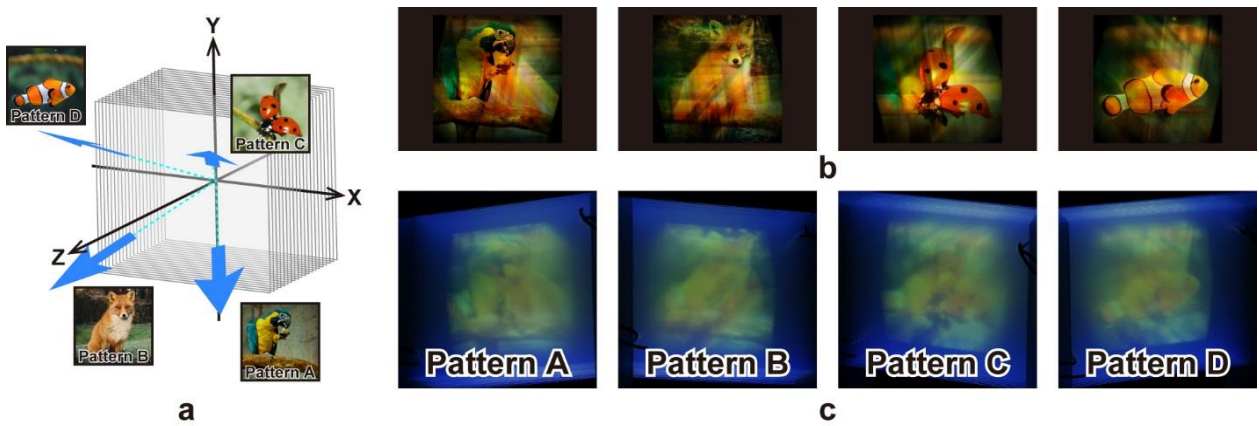

**Figure 4.** Prototype of the volumetric display projecting four full-colour patterns. (a) Scheme of the prototype that indicates the projection directions of the patterns. (b) Projected patterns from different viewpoints of the 3D structure in the simulation. (c) Projected patterns of the volumetric display. The photographs used as the original images can be found at http://free-photos.gatag.net/?s=201304180900 (pattern A), https://unsplash.com/photos/HQqIOc8oYro (patterns B), https://unsplash.com/photos/clkYWlgOHIQ (pattern C) and https://www.pexels.com/photo/clown-fish-swimming-128756/ (pattern D). All of them are licensed under the Creative Commons Public domain (https://creativecommons.org/publicdomain/zero/1.0/deed.en).

## Experimental evaluation

The quality of the 3D images of the proposed volumetric display partially depends on the number of the films since the transparency of the films used in practice is not perfect. This type of number-of-layer dependency of the image quality is evaluated by performing an experiment based on the setup, as schematically shown in Fig. 5a. A single layer on which a 2D pattern comprising red, green and blue circles is printed and is sandwiched by multiple layers of transparent films (namely, films without printed patterns). Specifically, $N_{UV}$ and $N_{VIS}$ denote the numbers of transparent films placed between the pattern-printed film and the ultraviolet light source and between the pattern-printed film and the camera, respectively. The increase in $N_{UV}$ and $N_{VIS}$ is unavoidable to realise a high resolution volumetric display in depth direction. However, the increase in $N_{UV}$ and $N_{VIS}$ further attenuates the ultraviolet light required to excite the florescent ink and the colour visible light emitted, respectively.

Figure 5b shows the captured images when $N_{VIS} = 0$ and $N_{UV} = 0, 5, ..., 25$. It should be noted that the camera was focused on the printed film. It is confirmed that the increase in $N_{UV}$ led to a significant decrease in the brightness and contrast of the image. This result is attributed to the transmittance of a film in the ultraviolet region (365 nm) that corresponds to 82 %. When $N_{UV} = 20$ and $N_{UV} = 25$, the ultraviolet light for excitation is attenuated to 2% and less than 1%, respectively. The images shown in Fig. 5c depict the captured images when $N_{UV} = 0$ and $N_{Vis} = 0, 5, ..., 25$. The decrease in the brightness and contrast of the image is lower than those of the results shown in Fig. 5b. This is because the transmittance of a film in the visible region is higher (about 90 %) than that in the ultraviolet region. In contrast, it is confirmed that the increase in $N_{Vis}$ leads to the deterioration of the images with respect to sharpness. This result is attributed to the diffusion and the refraction of the emitted visible light at the film surfaces. The aforementioned findings indicate that the transparency of the mother materials for inkjet printing radically affects the quality of the volumetric display. This is an important result that can be explored in detail in future studies.

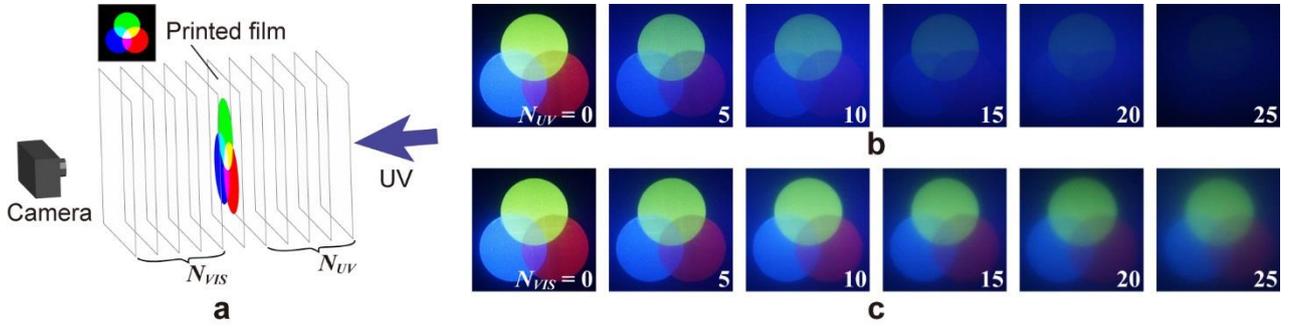

**Figure 5.** Evaluation of an effect based on a number of films. (a) Experimental setup. Captured images based on the number of clear films placed (b) between the printed film and the ultraviolet source and (c) between the printed film and the camera.

## Discussion

This section discusses the resolution of the proposed volumetric display. As described in the demonstration of a full-colour 3D image representation, the inkjet printer used in this study can print 2D patterns of fluorescent inks at an in-plane high resolution (a maximum of 5,760 × 1,440 dpi). However, the depth resolution (out-of-plane resolution) is only approximately 42 dpi, as determined by the film thickness (0.1 mm) and gaps (0.5 mm) between the films. In addition, the number of the films corresponds to only 20. Future studies will examine further improvements in the resolution of the display by increasing the number of the films. With respect to these films, increased transparency is indispensable to avoid deterioration of the image with the increase of the films (as quantitatively evaluated previously). For example, a heat-curable resin polydimethylsiloxane (PDMS) is used in various optical applications [18–20] due to its superior transparency. The transmittance of the PDMS exceeds 90% when the optical path length corresponds to 10 mm; thus, the transmittance of a 0.1 mm thick, PDMS film could exceed 99%.

Next, the projected full-colour patterns of the 3D structure designed by the extended algorithm are discussed. As shown in Figs. 3 and 4, it is confirmed that the algorithm can be successfully adapted to the full-colour images. However, certain background noises were observed in the computer simulations. It is considered that such a background noise can be suppressed by introducing an iteration algorithm proposed in a previous study [15] or by other optimisation methods. The projected patterns of the experimentally fabricated devices deteriorated when compared with the simulation results. It is presumed that the deterioration is caused by the degradation of 2D patterns printed on the film with respect to the contrast (brightness) as well as the blurring effects, as observed in Fig. 5c. The degradation of the contrast of the patterns is attributed to the absorption of the emitted light by the films. Thus, the contrast can be decreased by using highly transparent materials as films. The blurring is minimised by increasing the display resolution in the depth direction, and this will also be examined in detail in a future study.

In addition to the image quality improvement, the aim of the study involved realising a dynamic volumetric display system for practical applications of the proposed algorithm. In a previous study, an optically addressing method was proposed based on photochromic materials, i.e., photoreactive materials with unique characteristics of a reversible colour transformation [21]. The volumetric display is developed using this method, and the algorithm experimentally demonstrated in this study is applied on the display.

In summary, in this study, a manufacturing method of a high resolution volumetric display based on inkjet printing is proposed and experimentally demonstrated. The inkjet printer enables the minute and automatic location of photoreactive materials (that are used as the voxels of the 3D images) at appropriate places. Moreover, an algorithm to design 3D structures is developed to project multiple full-colour patterns, and volumetric displays based on the proposed method are experimentally demonstrated. Quantitative analysis of the observed images is performed to clarify important future research agenda.

# Methods

## Volume construction

The quantum yield of red, green and blue inks corresponds to 0.43, 0.85 and 0.89. Furthermore, 0.5 mm thick acrylic plates are used as spacers and placed between the twenty films.

A method to create the twenty cross-sectional images of a 3D object is described, as shown in Fig. 2b. The object data a created with computer graphics software and is composed of 3D position coordinates $(x, y$ and $z)$ and bright information $(R, G, B)$. It is assumed that the twenty films are placed on $z = Z_1, \dots, Z_{20}$, and a point of the 3D object is at $(O_x, O_y, O_z)$. When $Z_n < O_z < Z_{n+1}$, the point is printed at $(O_x, O_y)$ of the film placed at $z = Z_n$. All the points are printed on the films in the same manner.

## Algorithm

The algorithm to create the cross-sectional images of the 3D structure that exhibits multiple 2D patterns is described. It is based on a previous study [13] and is incremented to correspond to the full-colour representation. The algorithm can design a 3D structure that exhibits an arbitrary number of patterns (there is a trade-off between the number of the patterns and their image quality). Nevertheless, for the purpose of simplicity, the case wherein the number of patterns corresponds to three is considered, as shown in Fig. 6a. Specifically, $V(x, y, z)$ is a voxel value of the 3D structure that indicates the amount of inks at $(x, y, z)$. In the study, the full-colour images are treated; thus, each of the voxel values comprise three colour components, namely red, green and blue that correspond to $V_R(x, y, z)$, $V_G(x, y, z)$ and $V_B(x, y, z)$, respectively.

The voxel value $V(x,y,z)$ can be determined as follows:

1. Each of the original patterns is set up on the direction in which it is required to be projected.
2. Perpendicular lines $P_A$, $P_B$ and $P_C$ are drawn from the voxel to the patterns A, B and C, respectively. These lines are referred to as projection axes of the patterns.
3. The voxel value $V(x,y,z)$ is calculated as shown in Eq. (1) wherein $a(u_a, v_a)$, $b(u_b, v_b)$ and $c(u_c, v_c)$ correspond to the pixel values of the original patterns A, B and C at the intersections with each projection axis. Each pixel consists of red, green and blue components as follows:

$$V(x,y,z) = a(u_a, v_a) \cdot b(u_b, v_b) \cdot c(u_c, v_c). \qquad (1)$$

Next, the projected patterns of the 3D structure comprising the voxels determined by Eq. (1) are considered. It is assumed that the pixel values of the projected patterns are given by summations of the voxel values along their projection axes, as shown in Fig. 6b. For example, $a'(u_a, v_a)$ corresponds to the pixel value of the projected pattern A and is given by Eq. (2) as follows:

$$a'(u_a, v_a) = \sum_{P_A} V(x,y,z). \qquad (2)$$

In addition, $a(u_a, v_a)$ is constant along the projection axes $P_A$, and thus $a'(u_a, v_a)$ can be represented as Eq. (3) as follows:

$$a'(u_a, v_a) = a(u_a, v_a) \sum_{P_A} b(u_b, v_b) \cdot c(u_c, v_c). \qquad (3)$$

As a result, the projected pattern is given by multiplying the original pattern and a background noise, which corresponds to interference from the other two patterns. The component of the original pattern in Eq. (3) tends to be more dominant than that of the background noise when 2D images are generally used as the original patterns. Therefore, the projected patterns are recognised as the original patterns used to determine the voxel values. The same trend is observed for the pixels of patterns B and C; therefore, three patterns are recognised from the 3D structure.

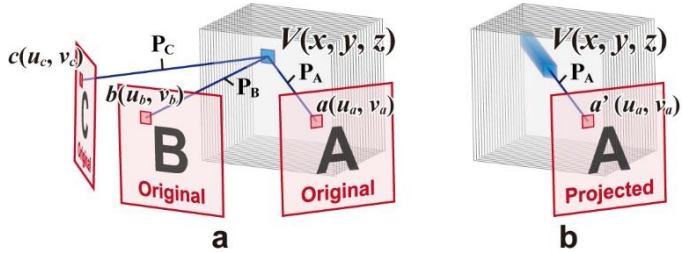

**Figure 6.** Schematic of the algorithm. (a) Voxel values are determined by multiplying the pixel values of the original patterns. (b) Projected pattern is given by the summation of the voxel values.

## Acknowledgements

This study was partially supported by the Japan Society for the Promotion of Science Grant-in-Aid No. 15J07684 and No. 25240015 and the Core-to-Core Program, A. Advanced Research Networks.

## Author contributions statement

R.H., Tomotaka S., M.N. and T.I. directed the project. Tomoyoshi S. suggested the basic concept of this study. H.N., A.S. and T.I suggested the algorithm demonstrated in this study. R.H., Tomotaka S. and T.I. designed and performed the experiments. R.H. Tomotaka S., N.M. and T.K. analysed the data. All authors contributed the discussions and reviewed the manuscript.

## Additional information

The authors declare no competing financial interests.

## Figure Legends

**Figure 1.** Concept of the study. (a) 3D glass structure projecting four monochromatic patterns in different directions. (b) Volume construction method enabling a high resolution. (c) Inkjet printing-based volumetric display projecting multiple full-colour 2D patterns.

**Figure 2.** Prototype of the volumetric display based on inkjet printing. (a) Overview under natural light. (b) Cross-sectional images printed on the films. (c) Original 3D image represented with computer graphics from three different viewpoints. (d) Photographs of the volumetric display when excited by ultraviolet light.

**Figure 3.** Prototypes of the volumetric display projecting three full-colour patterns. (a) Cross-sectional images printed on the films. (b) Original images recorded on the volume. (c) Projected patterns from different viewpoints of the 3D structure in the simulation. (d) Projected patterns of the volumetric display. The photographs used as the original images can be found at http://free-photos.gatag.net/?s=201304180900 (pattern A), https://unsplash.com/photos/HQqIOc8oYro (patterns B) and https://unsplash.com/photos/clkYWlgOHIQ (pattern C). All of them are licensed under the Creative Commons Public domain (https://creativecommons.org/publicdomain/zero/1.0/deed.en).

**Figure 4.** Prototype of the volumetric display projecting four full-colour patterns. (a) Scheme of the prototype that indicates the projection directions of the patterns. (b) Projected patterns from different viewpoints of the 3D structure in the simulation. (c) Projected patterns of the volumetric display. The photographs used as the original images can be found at http://free-photos.gatag.net/?s=201304180900 (pattern A), https://unsplash.com/photos/HQqIOc8oYro (patterns B), https://unsplash.com/photos/clkYWlgOHIQ (pattern C) and https://www.pexels.com/photo/clown-fish-swimming-128756/ (pattern D). All of them are licensed under the Creative Commons Public domain (https://creativecommons.org/publicdomain/zero/1.0/deed.en).

**Figure 5.** Evaluation of an effect based on a number of films. (a) Experimental setup. Captured images based on the number of clear films placed (b) between the printed film and the ultraviolet source and (c) between the printed film and the camera.